\pdfoutput=1
\documentclass[10pt, aps,pre, twocolumn, groupedaddress]{revtex4-2}

\usepackage{easyReview}

\usepackage{babel}
\usepackage{lipsum}
\usepackage{mathtools}
\usepackage{graphicx}
\usepackage{dcolumn}
\usepackage{amsmath}    
\usepackage{amssymb}
\usepackage{bm}

\usepackage{latexsym}
\usepackage{verbatim}
\usepackage{color}
\usepackage[caption=false]{subfig}
\usepackage{epstopdf}

\usepackage{easyReview}

\def\beq{\begin{equation}}
\def\eeq{\end{equation}}

\def\beq{\begin{equation}}                          
\def\eeq{\end{equation}}                          
\def\bea{\begin{eqnarray}}                          
\def\eea{\end{eqnarray}}

\DeclareRobustCommand{\uvec}[1]{{%
  \ifcsname uvec#1\endcsname
     \csname uvec#1\endcsname
   \else
    \bm{\hat{\mathbf{#1}}}%
   \fi
}}

\newcommand{\sk}[1]{\textcolor{black}{#1}} 
\newcommand{\mk}[1]{\textcolor{black}{#1}}
                   
\preprint{}

\usepackage{hyperref}

\begin{document}

\title{Defect Localization by Vanishing Deviatoric Stress in Active Nematics} 

\author{Sameer Kumar}
\email[]{sameerk@iitk.ac.in}
\affiliation{Department of Physics, Indian Institute of Technology Kanpur, Kanpur, India - 208016}
\author{Manas Khan}
\email[]{mkhan@iitk.ac.in}
\affiliation{Department of Physics, Indian Institute of Technology Kanpur, Kanpur, India - 208016}


\begin{abstract}
Collective stress generation in cellular monolayers is a key phenomenological process governing coordinated migration and emergent multicellular dynamics. We employ a generic active nematics model to investigate stress generation and its associated properties. By analyzing the maximal principal stress and its correlation with the nematic director across different activity strengths, we find that the principal stress aligns perpendicular (parallel) to the nematic director for extensile (contractile) activity. In the turbulent regime, we identify a rotation-invariant scalar measure of the in-plane deviatoric stress whose zero-level contour coincides with the locations of all $\pm 1/2$ topological defects (both nematic and \mk{principal} stress defects) are localized. This feature is robust and remains unchanged with variations in both the magnitude and nature (extensile or contractile) of activity. Our findings {thus open up} a new route to probe the \mk{spatial alignment from the} mechanical and rheological properties of confluent cell layers, where stress measurements are more accessible than detailed cell shape or size characterization.
\end{abstract}

\maketitle

\section{Introduction}

Active systems, intrinsically out of equilibrium, show interesting behavior such as pattern formation, coherent motion and the spatiotemporal changes \cite{MarchettiRMP2013, CatesARCMP2015, RamaswamyJStatMech2017, ProstNatPhys2015, KumarPhysRevLett2025}. Active systems are found at various length scales such as micro-organisms like cytoskeletal filaments \cite{doostmohammadi2016}, bacteria colonies\cite{CopenhagenNatPhys2021}, cells in tissue \cite{FriedlNatPhys2009, Saw2017}, and macro-organisms like fish schools \cite{VicsekPRL1995}, bird flocks \cite{TonerPRL1995}, etc. 
In essential physiological functions including morphogenesis, wound healing, and tissue regeneration, the prevalent mode of cellular migration is collective \cite{FriedlNatRevMolCell2009, RorthARCDB2009, Saw2017, SerraPicamal2012, GiampieriNCB2009}. Collective cellular migration plays a role not only in development, physiology, and repair, but also in devastating diseases including cancer \cite{FriedlIJDB2004, LiLIBT2013}. 
These emergent \mk{phenomena} are found to be due to the collective cell guidance by cooperative forces \mk{where} the local cell migration follow local orientation of maximal principal stress \cite{TrepatTCB2011, SerraPicamal2012, TambeNatMat2011Nature}.\\

Recent studies have shown that theories of liquid crystal hydrodynamics—particularly active nematics—successfully capture many of the collective behaviors observed in confluent cell layers \cite{SimhaRamaswamyPRL2002, ThampiPRL2013, DoostmohammadiNatComm2016, Doostmohammadi2018, ThijssenPNAS2021, SameerPRE2020, KumarPhysRevE2022, NejadPRE2025}. The motivation for using the active nematic framework arises from the fact that individual cells often become anisotropic while moving, tending to migrate along their elongated axis \cite{SimhaRamaswamyPRL2002, Duclos2017, LiPNAS2017}. When many such elongated cells align parallel to one another, they form an ordered state known as a nematic phase \cite{MarchettiRMP2013}, as observed in cell types such as fibroblasts \cite{Duclos2017} and human mesothelial cell line LP-9 cells \cite{ZhangAPL2021, NejadNatComm2024}. Some cell types, including Madin–Darby Canine Kidney (MDCK) cell lines \cite{TambeNatMat2011Nature, ComellesELife2021, Saw2017, AscioneJRSI2023, NejadNatComm2024}, remain nearly isotropic in shape when stationary. Despite this, these systems can still exhibit a remarkable phenomenon known as activity-driven turbulence at  low Reynolds numbers, commonly referred to as active turbulence \cite{DombrowskiPRL2004, DoostmohammadiNatComm2016, Doostmohammadi2018}. In a fully developed active turbulent state, the system continuously generates and annihilates pairs of topological defects with charges $+1/2$ and $-1/2$ \cite{MarchettiRMP2013, ThampiPRL2013, ThampiEPL2014, Saw2017}. This dynamic process is accompanied by chaotic cellular flows, which are driven by regions of high vorticity within the cell layer \cite{BlanchMercader2018, Duclos2017}.\\

To gain deeper insight into the behavior of active systems, it is essential to examine how collective forces are generated at the particle or cellular level. In particular, the relationship between topological defects arising from cellular anisotropy and possible singularities in the underlying stress field \mk{still remain an intriguing question}. \sk{In the present work, we focus on the principal stresses
generated through interactions among neighboring cells
(or particles)  and its relation to the local orientation of the cells.}. This perspective is especially powerful in biological systems, where measuring stresses and their components through traction force microscopy (TFM) provides richer mechanical information than analyses based solely on cell shape elongation \mk{and their orientational correlations, i.e.,} the nematic order parameter \cite{TambeNatMat2011Nature, DoxzenPNAS2012, MamblonaBiophysJorn2014, MuellerPRX2020, NejadNatComm2024, TrepatTCB2011, ZimmermannBiophysJ2014, SaraswathibhatlaPRX2020}. Experimental observations indicate that motile cells often tend to align and migrate along the direction of maximal local stress \cite{TambeNatMat2011Nature, TrepatTCB2011, SerraPicamal2012}. {However,} some studies report a strong correspondence between the principal stress axis and the cell elongation axis with notable  \mk{(misalignment)} between these directions \cite{NejadNatComm2024, rozman2025why}.  Recent works on confluent cell layers further show that the distribution of traction forces reveals complex mechanical and viscoelastic properties of cellular tissues \cite{TambeNatMat2011Nature, TrepatTCB2011}. This raises important questions — how does the cooperative motion of cells and their alignment with the maximal principal stress depend on key parameters such as active force strength and substrate friction? Can the components of the stress tensor (or the principal stress) give rise to geometrical singularities or defect-like structures analogous to those defined by the nematic director field? If yes, then what additional features do they {add to} the stress defects? Exploring the spatial organization and dynamics of principal stresses may therefore offer a new framework for understanding the complex flow { patterns} and turbulent state  characteristics of active nematic systems.\\

We adopt the generic continuum model of active nematics \cite{SimhaRamaswamyPRL2002, ThampiPRL2013, DoostmohammadiNatComm2016, Doostmohammadi2018, ThijssenPNAS2021} to study the system. The \mk{corresponding} hydrodynamic equations describe active nematic particles suspended in an incompressible fluid. The orientational order of the particles is represented by the nematic order parameter tensor ${Q}_{ij}$, while the combined velocity field of the active units and the surrounding fluid is denoted by $\mathbf{v}$.
From numerical simulations, we obtain several key observations, first, the maximal principal stress and nematic director exhibit a systematic mutual alignment; they align perpendicularly in the case of extensile activity, whereas a parallel alignment is observed for contractile activity. Second, the principal stress orientation field forms topological defects of charge $+1/2$ and $-1/2$, identical to those observed in the nematic director field. Although the topology of the defects is the same, the effect of activity on the stress defects differs from that on the nematic defects. Finally, in the active turbulent regime, we identify a distinct isoline defined by in-plane deviatoric stress components that acts as an organizing backbone for the system, along which all the $\pm1/2$ topological defects (both nematic and maximal principal stress defects) are localized. Remarkably, this feature persists across varying activity strengths and is independent of the extensile or contractile nature of the \mk{stress}. Experimentally, defect localization is observed  in \cite{HeadNatPhy2024} where $+1/2$ nematic defects are appears on zero-line defined by vorticity and the strain rate tensor.\\


\section{Model}
\label{model}
We describe the system as  two dimensional active nematics suspended in an incompressible fluid.  It's dynamics are described in terms of the combined velocity of \mk{the} fluid and active particles, ${\bf v(r},t)$\mk{,} and the nematic order parameter ${Q}_{ij}=S(n_i n_j-\frac{1}{2}\delta_{ij})$, where  $S$ is the magnitude of the nematic order and ${\bf n}_i$ is the unit director ($i=1,2$ in two dimensions). The equations of motion are adopted from the liquid crystal hydrodynamics \mk{by adding} an active term \mk{that is} proportional to $Q_{ij}$ such that \mk{the} gradient in $Q_{ij}$ produces a flow in the system. Evolution of $Q_{ij}$ and the momentum $\rho {\bf v}$, together with the incompressibility condition $\nabla \cdot {\bf v}=0$, is given by \cite{BerisandEdward1995, PGdeGenneBook1995},

\begin{equation}
\centerline{$ (\partial_t + v_k\partial_k){Q}_{ij} -S_{ij}=\Gamma{H}_{ij},$}
\label{eq: 1}
\end{equation}
and,
\begin{equation}
\centerline{$\rho(\partial_t + v_k\partial_k)v_i = \partial_j \sigma_{ij},$}
\label{eq: 3}
\end{equation}

where,  $S_{ij}=(\xi E_{ik}+\Omega_{ik})(Q_{kj}+\delta_{kj}/2)+(Q_{ik}+\delta_{ik}/2)(\xi E_{kj}-\Omega_{kj}) - 2 \xi (Q_{ij}+\delta_{ij}/2)Q_{kl}\partial_k v_l$ is the generalized advection term described in terms of the strain rate tensor, $E_{ij}=(\partial_i v_j+\partial_j v_i)/2$ and vorticity tensor $\Omega_{ij}=(\partial_i v_j-\partial_j v_i)/2$,  $\xi$ {being} the flow alignment parameter. Rotational diffusivity is denoted by $\Gamma$, and the relaxational dynamics of the system is embodies by the  molecular field, $H_{ij}=-\frac{\delta F}{\delta Q_{ij}}$, determined from the free energy, $F=\int dA[{C} (1 -Q_{ij}Q_{ij})^2 +\frac{L}{2} \vert \nabla {\mathcal {Q}} \vert^2]$, where $C$ sets the energy scale for Landau-de Gennes free energy \cite{PGdeGenneBook1995} and $L$ is the Frank elastic constant within the single elastic
constant approximation \cite{FrankRSC1958}.   

Total hydrodynamic stress is given by the stress tensor, $\sigma_{ij}=\sigma_{ij}^{\mathrm{p} }+\sigma_{ij}^{\mathrm{v}}+\sigma_{ij}^{\mathrm{a}}$. Where $\sigma_{ij}^{\mathrm{p}}=-p \delta_{ij}+2\xi(Q_{ij}+\delta_{ij}/2)Q_{lk}H_{kl}-\xi H_{ik}(Q_{kj}+\delta_{kj}/2)-\xi (Q_{ik}+\delta_{ik}/2)H_{ik}-\partial_iQ_{kl}\frac{\delta F}{\delta \partial_j Q_{lk}}+Q_{ik}H_{kj}-H_{ik}Q_{kj}$ is the passive stress with $p$ as the isotropic pressure, $\sigma_{ij}^{\mathrm{v}}=-2\eta E_{ij}$  is viscous stress,  and $\sigma_{ij}^{\mathrm{a}}=-\zeta Q_{ij}$ is the active stress \cite{BerisandEdward1995, SimhaRamaswamyPRL2002, CatesPRL2008, FieldingPRE2011}, with $\zeta$ as the strength of activity. 

We simulate the coupled system, \mk{defined by} Eq. (\ref{eq: 1}) and Eq. (\ref{eq: 3}), using hybrid Lattice-Boltzmann algorithm \cite{DennistonPRE2001, MarenduzzoPRE2007, FieldingPRE2011} with $\rho=200$ and $\eta=1/6$ in lattice Boltzmann unit such that the Reynold's number $R_e << 1$. The dimensions of other relevant quantities e.g., $L$ and $\xi$ are $[ML^2T^{-2}]$ and $[1]$, respectively. The microscopic coherent length is given by, $L_0=\sqrt{L/C}$.  In the simulations, we take $\xi=0.3, \Gamma=0.1, \ L=0.5, \ C=0.1$ (unless specified). We study the system \mk{at varying strength of} activity, $\zeta$  for system size, $= 256$ with periodic boundary condition. 

\begin{figure*}      
{\includegraphics[width=14cm]{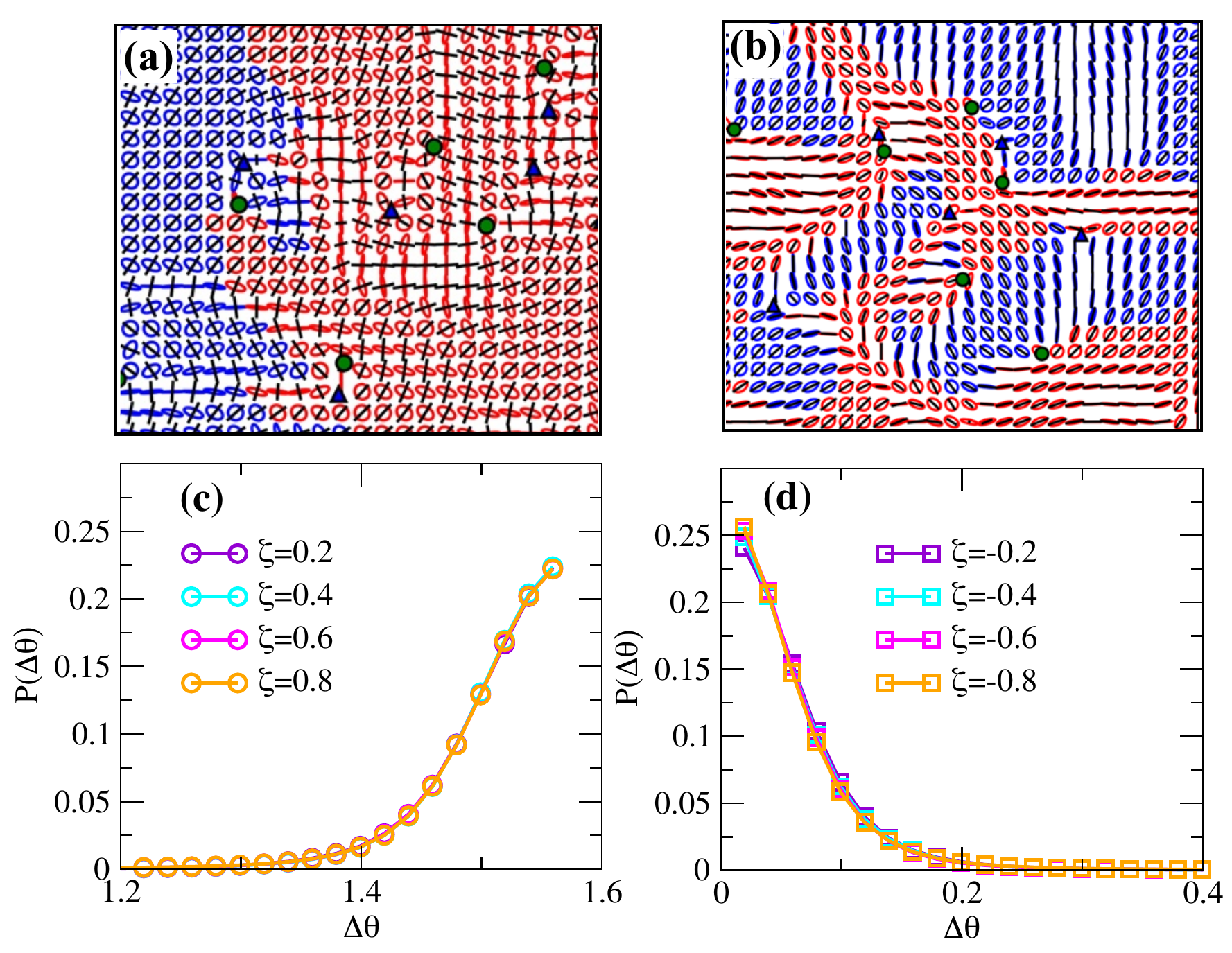}} 
\caption{Snapshots from the simulation \mk{exhibiting principal stress and nematic director field} for \mk{activity} $\zeta=0.8$ (a), and $\zeta=-0.8$ (b). Ellipses represent the principal stress where the major (minor) axis of an ellipse corresponds to the local maximal principal stress, $\sigma^p_{max}$ (minimum, $\sigma^p_{min}$) and{, therefore,} circles corresponds to {the} cases when $\sigma^p_{max}=\sigma^p_{min}$. Color of the ellipses are red (blue) when  $\sigma_{iso}>0$ ($\sigma_{iso}<0$). Black rods show  the local nematic director, plotted on top of the ellipses. Filled red circles and blue triangles represents $+1/2$ and $-1/2$ nematic defects, respectively. \mk{Probability density} of  $\Delta \theta$, where $\Delta \theta$ is the angle difference between orientations of nematic director ${\bf n}$ and the maximum principal stress director ${\bf n}_{s}$\mk{, are shown} for $\zeta > 0$ (c) and $\zeta<0$ (d), for different values of $\zeta$. Here the angle difference $\Delta \theta$ is in radians.}
\label{fig:101}
\end{figure*}

\section{Results}
\label{results}
We start with a uniform initial configuration \sk{where the local nematic directors, ${\bf n}$ are aligned in the x-direction} and let the system evolve to the  active turbulent state \sk{where continuous creation and annihilation of the $\pm 1/2$ defects occurs. Further details are provide in the appendix \ref{Appendix1}}\cite{ThampiPRL2013, ThampiEPL2014, DoostmohammadiNatComm2016}. \mk{Typical} snapshots from the simulation \mk{describing turbulent states of the system are shown} in Fig. \ref{fig:101}. Topological defects corresponding to the nematic director, ${\bf n}$, represented by the black lines. We also show the \mk{spatial variation of the} magnitude of the maximal and the minimum principal stresses, $\sigma_{max}$ and $\sigma_{min}${, respectively, by representing them as the major and minor axes of} ellipses. In general, the maximal principal stress, ${\sigma_{max}}$ at a point in the space aligns to the local normal stress{, represented by,} $\sigma_{iso} = (\sigma_{xx}+\sigma_{yy})/2${,} where the shear stress is zero, i.e., $\sigma_{xy}=0$  \cite{MalvernBook1969, GurtinBook1981}. 
\sk{Furthermore, we calculate the orientation of the maximum principal stress, $\theta_p=\frac{1} {2}\tan^{-1}\left(\frac{2\sigma_{xy}}{\sigma_{xx}-\sigma_{yy}}\right)$. The corresponding principal stress direction is represented by the unit vector $\mathbf{n}_{p}=\left(\cos 2\theta_p,\sin 2\theta_p\right)$. The maximum and minimum principal stresses are then expressed as $\sigma_{\max}=\sigma_{\mathrm{iso}}+J,\qquad
\sigma_{\min}=\sigma_{\mathrm{iso}}-J$, where $J=\sqrt{\left(\frac{\sigma_{xx}-\sigma_{yy}}{2}\right)^2+\sigma_{xy}^{,2}}$ is the magnitude of the in-plane deviatoric stress. Equivalently, $J$ is the radius of Mohr's circle and equals one-half of the principal stress difference, i.e., $J=(\sigma_{\max}-\sigma_{\min})/2$. Since, $J$, depends only on the stress invariants, it is invariant under coordinate rotations.} Now, we discuss the results in detail: 


\subsection{Alignment of the maximal principal stress} 
\mk{In} the snapshots from the simulation\mk{, (fig. \ref{fig:101})} {the} principal stresses are represented by ellipses and nematic directors are {shown} as rods. In the fully developed turbulent state (see the Appendix \ref{Appendix1} for details), nematic director\mk{s} are aligned perpendicular to the major axis of the ellipses\mk{, i.e., the}  axis of maximal principal stress, $\sigma_{max}$ {for} $\zeta=0.8$, see fig. \ref{fig:101}(a). In contrast\mk{, the} nematic directors are aligned parallel to the major axis {for} $\zeta=-0.8$, see fig. \ref{fig:101}(b). It is also reported experimentally that in a cellular mono-layer, cells migrate in the direction of maximal normal stress and lowest shear stress \cite{TambeNatMat2011Nature}. \mk{This signifies that} the director in  active nematics should show a clear alignment\mk{, which is observed in our simulations}. The $\pm 1/2$ topological defects are formed due the bend deformation, representing the turbulent state with extensile stress ($\zeta>0$, Fig. \ref{fig:101}(a) ) \cite{ThampiEPL2014, ThampiPRL2013, DoostmohammadiNatComm2016}. These defects increases with an increase in the strength of activity ($\zeta >0$) \mk{(Please note that the number of defects is similar for positive and negative value of $\zeta$ as shown in panel (a) and (b))}. We also observe  that the majority of the $\pm 1/2$ defects are in the region of tension (i.e., when $\sigma_{iso}>0$, red ellipses) or equivalently, more defects leads to {higher} tension. \\

To quantify the alignment of the maximal principal stress, $\sigma_{max}$, we  calculate the probability distribution of angle difference between the orientation of  nematic director  ${\bf n}$ and the orientation of the maximum principal stress director  ${\bf n}_p$ for different activity strength  $\zeta$. We observe that the principal stress aligns perpendicular (parallel) to the nematic director for extensile (contractile) system.  {S}imilar results have been observed for the MDCK monolayer \cite{ NejadNatComm2024}\mk{,} where separate clusters of the contractile and extensile regions are found in the same monolayer.  Fig. \ref{fig:101}(c,d) show the  {variation in} $P(\Delta \theta)$  for extensile activity ($\zeta>0$, Fig. \ref{fig:101}(c)) and contractile activity ($\zeta<0$, Fig. \ref{fig:101}(d)).  {T}he distribution shows single peak, at $\Delta \theta = \pi/2$ for $\zeta >0$ and $\Delta \theta = 0$ for $\zeta <0$.  There is no significant change in the height of the peak if we change the value of $\zeta$ (\mk{when $\zeta > 0$}) but see moderate narrowing of the peaks by increasing \mk{(decreasing?)} the strength \mk{of} $\zeta$ \mk{(when $\zeta < 0$)}. The appearance of one  peak suggest that the principal stress always align perpendicular (parallel) to the nematic director for \mk{an }extensile (contractile) system. The appearance of peak at $\pi/2$ (or $0$) can be understood by looking at  {the definition of} the active stress, $\sigma^a_{ij} = -\zeta Q_{ij}$, which is the only dominating stress in the system.\\

\subsection{Cross correlation  between nematic and principal stress fields}
\begin{figure}      
{\includegraphics[width= \linewidth]{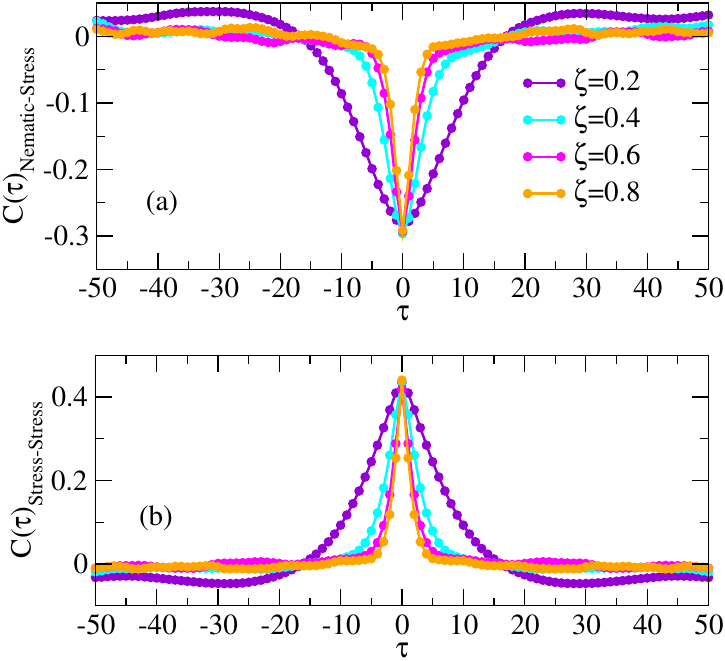}}  
\caption{Cross correlation function $C(\tau)$ plot against lag time $\tau$ for different strengths of activity, $\zeta>0$.  Cross correlation between nematic and the maximal principal stress directors, $C_{{\bf n},{\bf n}_s}$ (a);  Autocorrelation  function for the maximal principal stress directors, $C_{{\bf n}_s,{\bf n}_s}$ (b).}
\label{fig:117}
\end{figure}

We quantify the dynamical coupling between the nematic director (${\bf n}$) and the maximal principal stress director (${\bf n}_{s}$) by computing the Pearson's cross-correlation coefficient, $C_{X,Y}(\tau)=\frac{\langle(X(t)-\bar{X})(Y(t+\tau)-\bar{Y})\rangle}{\Delta_X \Delta_Y}$, where $X$ and $Y$ denote the relevant observables, and $\Delta_i$ is the corresponding standard deviation. Fig. \ref{fig:117}(a) shows $C_{{{\bf n}},{\bf n}_s}(\tau)$ as a function of lag time $\tau$ or different strengths of activity $\zeta (>0)$. At zero lag, the correlation is strongly negative, indicating a preferential perpendicular alignment between the nematic and stress directors, consistent with the orientation\mk{al} alignment {shown} in Fig. \ref{fig:101}. The magnitude of the correlation decreases with increasing $\tau$, reflecting the temporal decorrelation of the two fields. Notably, the decay becomes progressively faster with increasing activity strength,  $\zeta$, \mk{which} lead\mk{s} to achieve the  active turbulence state fast\mk{er} (see Appendix \ref{Appendix1} for details)\mk{, and} hence accelerates the reorientation dynamics of both ${\bf n}$ and ${\bf n}_s$.
In Fig. \ref{fig:117}(b), we show 
$C_{{{\bf n}_s},{\bf n}_s}(\tau)$, {i.e.,} the autocorrelation function  for the maximal principal stress ($\sigma_{max}$) director. As expected, the correlation is positive at $\tau=0$ and decays with increasing lag time. The decay rate increases with activity strength, exhibiting a trend consistent with that observed for the cross-correlation in Fig. \ref{fig:117}(a).

\subsection{Topological characteristics of nematic and  maximal principal stress defects}
\begin{figure*}      
{\includegraphics[width=12cm]{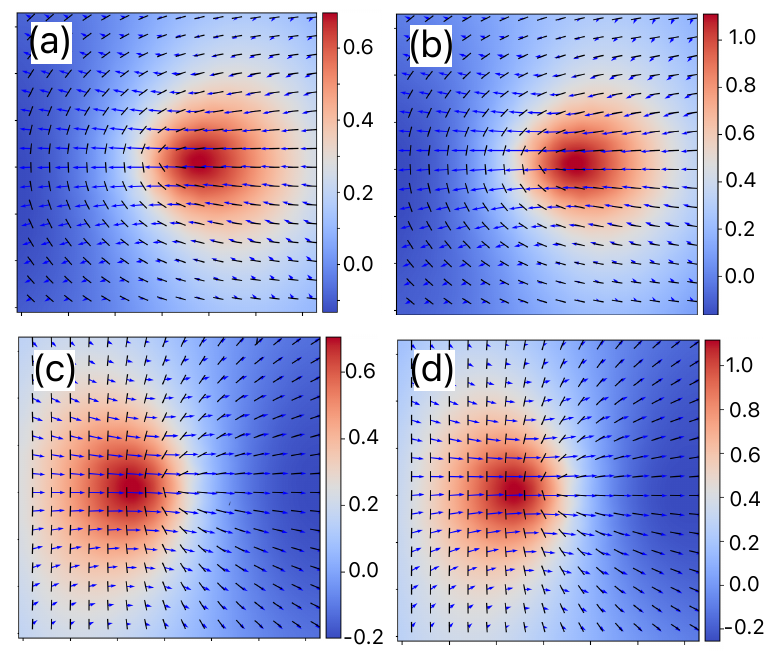}}
\caption{\mk{Nematic and maximal principal stress defects.} Average iso-stress \mk{field} $\sigma_{iso}^{avg}$ \mk{is} represented by  color \mk{variation according to the color-bar}, near {a} $+1/2$  (a,b) nematic and (c,d) maximal principal stress defect. \mk{Strength of activity increases from} {l}eft to right \mk{as} $\zeta= \ 0.5, \ 0.8$, respectively. Black lines show  the nematic director (a,b) and maximal principal stress director ${\bf n}_s$ (c,d). Blue quivers {exhibit} the local flow directions ${\bf v}_i$. We first calculate $\sigma_{iso,i}=\frac{\sigma_{xx}+\sigma_{yy}}{2}$ for each $+1/2$ defect, \mk{denoted as} $i$\mk{,} and then {obtain the} average iso-stress  {as} $\sigma_{iso}^{avg}= \frac{1}{N_{+1/2}}\sum_i^{N_{+1/2}} \sigma_{iso,i}$\mk{,} where $N_{+1/2}$ is the number of $+1/2$ defects. }
\label{fig:109a}
\end{figure*}

\sk{To get the average stress distribution and flow patterns near the principal stress defects,} we calculate the average iso-stress, $\sigma_{iso}^{avg}$, near the $+1/2 $ defect generated for the nematic director ${\bf { {n}}}$ and the maximal principal stress director ${\bf {{n}}}_s$ and \mk{show them} in Fig. \ref{fig:109a}. We find that  $\sigma_{iso}^{avg}$ and its \mk{spatial} distribution at the core of the nematic $+1/2$ defect and the maximal principal stress $+1/2$ defects {are} slightly different from each other. \sk{The region of maximum tension, where $\sigma_{iso}^{avg}>0$, is at the core of the $+1/2$ nematic defects (see fig. \ref{fig:109a}a,b) whereas in is away from the core of the $+1/2$ maximal principal stress defects.} 
This implies that the $\sigma_{max}$ defect core always appears \mk{at a distance from} where \mk{the} isotropic part of the total stress is maximum. This also suggests that the activity affects the stress distribution near the nematic and stress defect differently. \mk{Our results are in agreement with experimental results} \cite{NejadNatComm2024} where the nematic defects and stress defects \mk{are shown }separately for LP-9 cell line. \\


\begin{figure*}      
{\includegraphics[width=\linewidth]{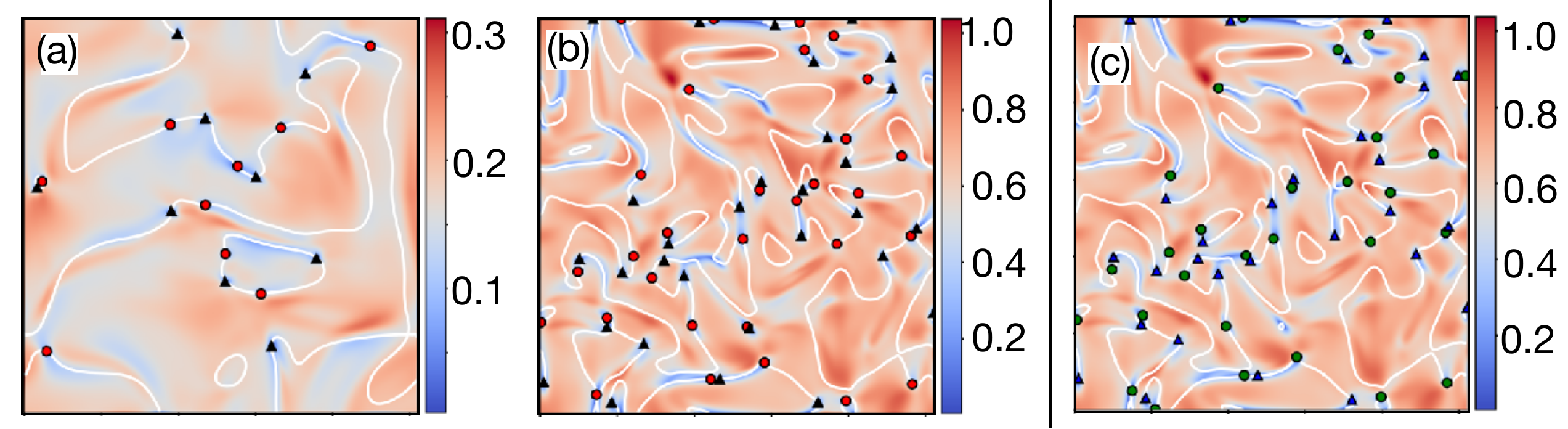}} 
\caption{\mk{Simulation} snapshots for defect localization in the turbulent state. \mk{The} color bar shows the magnitude of the in-plane deviatoric stress,  $J=\sqrt{(\frac{\sigma_{xx}-\sigma_{yy}}{2})^2-\sigma_{yx}^2}$.  $+1/2$ and $-1/2$ maximal principal stress defects are \mk{marked} by filled red circles and black triangles, respectively (a,b) and for nematic defects they are marked by filled green circles and blue triangles, respectively (c). The iso-lines (while color) represents the points where $\sigma_{xx}-\sigma_{yy}=0.$  The activity strength, $ \zeta = 0.2$ (a) and $\zeta=0.8$ (b,c).}
\label{fig:111}
\end{figure*}

\subsection{Defect localization}
In the fully developed active turbulent state, we calculate the in-plane deviatoric  stress $J$, which can be equivalently expressed as, $J = \frac{\vert \sigma_{max}-\sigma_{min} \vert}{2}$, and plot it as \mk{a} color map in Fig.  \ref{fig:111} for different strength of active stress, $\zeta$. We  \mk{mark} the maximal principal stress defects by filled red circles ($+1/2$-defects) and black triangles ($-1/2$-defects), Fig. \ref{fig:111}(a,b). Further, the nematic  defects by filled gree circles ($+1/2$-defects) and blue triangles ($-1/2$-defects), Fig. \ref{fig:111}(c).
The defect locations correspond to points where $J $ vanishes or attains a minimum. At these points, the stress tensor becomes locally isotropic and degenerate, implying that the principal directions are undefined. This is precisely the condition required for a topological defect in an orientation field. Consequently, principal stress defects \mk{appear} at minima of $J $ where the orientation of principal stress eigenvectors becomes singular.\\
Further, $J=0$ requires both $\sigma_{xx}-\sigma_{yy}=0$ and $\sigma_{xy}=0$. The condition $\sigma_{xx}-\sigma_{yy}=0$ defines an isoline along which the normal stresses are equal, while defect cores occur at locations where the shear stress simultaneously vanishes ($\sigma_{xy}\rightarrow 0$). Consequently, the stress tensor becomes locally isotropic ($J=0$) at every defect core, and the isoline $\sigma_{xx}-\sigma_{yy}=0$ provides a geometric backbone for defect localization. We demonstrate these features for activity strengths $\zeta=0.2$ and $0.8$ in Fig. \ref{fig:111}(a-c), respectively.
\\

This behavior is also observed for the nematic defects, as shown in Fig. \ref{fig:111}(c). To understand this correspondence, we recall that the active stress satisfies $\sigma_{ij}^{\mathrm{active}} \propto Q_{ij}$, and both tensors are symmetric and traceless. In two dimensions, the nematic tensor can be represented by the scalar pair $q=(S/2)\cos(2\theta)$ and $p=(S/2)\sin(2\theta)$, where $S=\sqrt{q^2+p^2}$ is the magnitude of the nematic order parameter. Although the total stress tensor is not strictly proportional to $Q_{ij}$, its deviatoric part has the same symmetric traceless structure. Motivated by this correspondence, we represent the deviatoric stress tensor ( $\sigma_{ij}^{\mathrm{dev}}$) by the analogous scalar pair $q_1=(\sigma_{xx}-\sigma_{yy})/2$ and $p_1=\sigma_{xy}$. The magnitude of the in-plane deviatoric stress is then given by $J=\sqrt{q_1^2+p_1^2}$, directly analogous to the definition of the nematic order parameter $S$. Consequently, the pair $(q_1,p_1)$ completely characterizes the deviatoric stress tensor and determines its principal directions in the same way that $(q,p)$ determines the nematic director.
\begin{figure}      
\centerline{{\includegraphics[width=8cm]{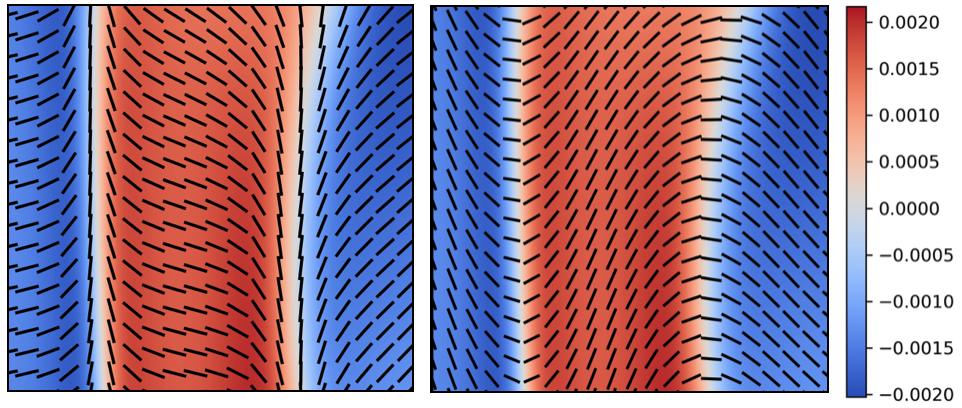}}}
\caption{At the onset of the active turbulence, plot of $\sigma_{ij}^{\mathrm{dev}}$ (left) directors and the $Q_{ij}$ directors (right). Color bar shows the magnitude of vorticity, $\omega=\mathbf{ \nabla \times v}$, for $\zeta=0.8$. }
\label{fig:120}
\end{figure}
We plot the directors corresponding to the deviatoric stress tensor, $\sigma_{ij}^{\mathrm{dev}}$, in Fig. \ref{fig:120} and observe that they closely resemble the nematic director field. In particular, the stress director field exhibits well-defined $\pm1/2$ defects whose locations coincide with those of the nematic defects. An additional feature emerges at the onset of the instability: under extensile activity ($\zeta>0$), the stress directors display a splay distortion, whereas the nematic directors exhibit the characteristic bend deformation. Consequently, the geometry of the stress $+1/2$ defects resembles that of nematic $+1/2$ defects in a contractile system, even though the underlying activity is extensile. As expected, the direction of motion of the stress $+1/2$ defects is identical to that of the maximum principal stress defects. More importantly, the coincidence of the stress and nematic defect locations follows from the fact that both defects occur where their respective deviatoric components vanish. Thus, although the stress and nematic fields are not identical, they share the same defect cores because the anisotropic part of the stress is predominantly generated by the active nematic order. These results remain valid for both type of active stresses, extensile ($\zeta>0$) \mk{and} contractile ($\zeta<0$). \\

We therefore conclude that the principal stress directions define a nematic-like field whose $\pm1/2$ defects coincide with the zero-level set of the in-plane deviatoric stress measure. Although the total stress tensor is not strictly proportional to the nematic tensor, its deviatoric part shares the same symmetric traceless structure. Consequently, the defect cores of the stress and nematic fields coincide because both correspond to the vanishing of their respective deviatoric components. Furthermore, the orientation of the maximum principal stress relative to the nematic director depends on the nature of the activity, aligning perpendicular to the director in extensile systems and parallel to it in contractile systems.

\section{Discussion}
\label{summery}
In this work we investigate  the mechanical stress field \mk{and associated defects} generated in an incompressible active nematic system by analyzing the principal stresses of the stress tensor rather than focusing solely on the nematic director field. This perspective allows us to directly connect the orientational property  of the stress field with the topology of defects in the active nematic turbulent regime. Key \mk{findings} of our study are  (i) the systematic relationship between the orientation of maximal principal stress and the nematic director and, (ii) existence of  a isoline derived from the in-plane deviatoric stress tensor where all the defects localizes. The maximal principal stress aligns perpendicular to the nematic director in extensile systems, whereas it aligns parallel to the director in contractile systems. This behavior follows naturally from the definition of the active stress tensor, which is proportional to the nematic order parameter tensor. In extensile systems, active stresses generate tensile forces along the axis perpendicular to the director due to outward pushing flows, while in contractile systems the stresses are concentrated along the director due to inward pulling forces. Our numerical results therefore provide a clear mechanical interpretation of how the nature of activity determines the orientation of principal stresses in active nematic flows.
When focusing on extensile active stress, we find that the principal stress directions form an orientational field with nematic symmetry and exhibits topological defects of charge $+1/2$ and $-1/2$, identical to the defects observed in the nematic director field. When we plot the average iso-stress $\sigma_{iso}^{avg}$, we find that region of tension appears at the core of the nematic defect, whereas it is away in the case of stress-defect. This \mk{indicates} that the   flow generated in the medium affect nematic and stress defects differently.

A striking observation from our analysis is that both the nematic and stress defects are localized on the isoline defined by $\sigma_{xx}-\sigma_{yy}=0$, which is derived from the in-plane deviatoric stress measure, $J$. Our results further show that the $\pm1/2$ defects of both the nematic and principal stress fields occur at locations where (J=0), corresponding to the complete vanishing of the in-plane deviatoric stress. At these points, the principal stresses become equal and the stress tensor is locally isotropic. The isoline $\sigma_{xx}-\sigma_{yy}=0$ therefore acts as a geometric backbone for defect localization, while the additional condition $\sigma_{xy}=0$ uniquely identifies the defect cores. This behavior is robust for both extensile and contractile activity and persists over a wide range of activity strengths. These findings establish a simple mechanical criterion for identifying stress defects in active nematic systems.\\

Such stress-based analysis of active turbulent state provide a complementary description of active nematic dynamics that focuses directly on mechanical quantities rather than on the cell shape elongation and its orientational order alone. From a physical perspective, this framework may help bridge theoretical models of active nematics with experimental measurements of stresses in cellular monolayers, where quantities such as principal stresses can be inferred using techniques like traction force microscopy and monolayer stress microscopy \cite{TambeNatMat2011Nature, Saw2017, NejadNatComm2024, TrepatTCB2011}.\\

More broadly, our findings suggest that the organization of stresses in active systems follows universal topological rules similar to those governing nematic order. In particular, the correspondence between nematic defects and principal stress defects indicates that mechanical stresses carry information about the underlying orientational dynamics. This insight may prove useful for interpreting experimental observations of collective cell behavior in epithelial tissues and other active materials, where stress fields and defect dynamics play a central role in driving large-scale flows and structural rearrangements. Further, it could be interesting to check these results in a system where the two different species are interacting collectively.\\

\section{Acknowledgment-}
SK thanks Amin Doostmohammadi for helpful discussion. The support and the resources provided by PARAM Sanganak under the National Supercomputing Mission, Government of India at the Indian Institute of Technology, Kanpur are gratefully acknowledged. \mk{MK thankfully acknowledges funding from Indian Institute of Technology Kanpur, and from SERB (presently ANRF), Govt. of India. through a core research grant (Grant No. CRG/2020/002723).}

\bibliographystyle{apsrev4-1}
\bibliography{references} 

\appendix
\section{End matter: Active turbulent state}
\label{Appendix1} 

\begin{figure}      
{\includegraphics[width=\linewidth]{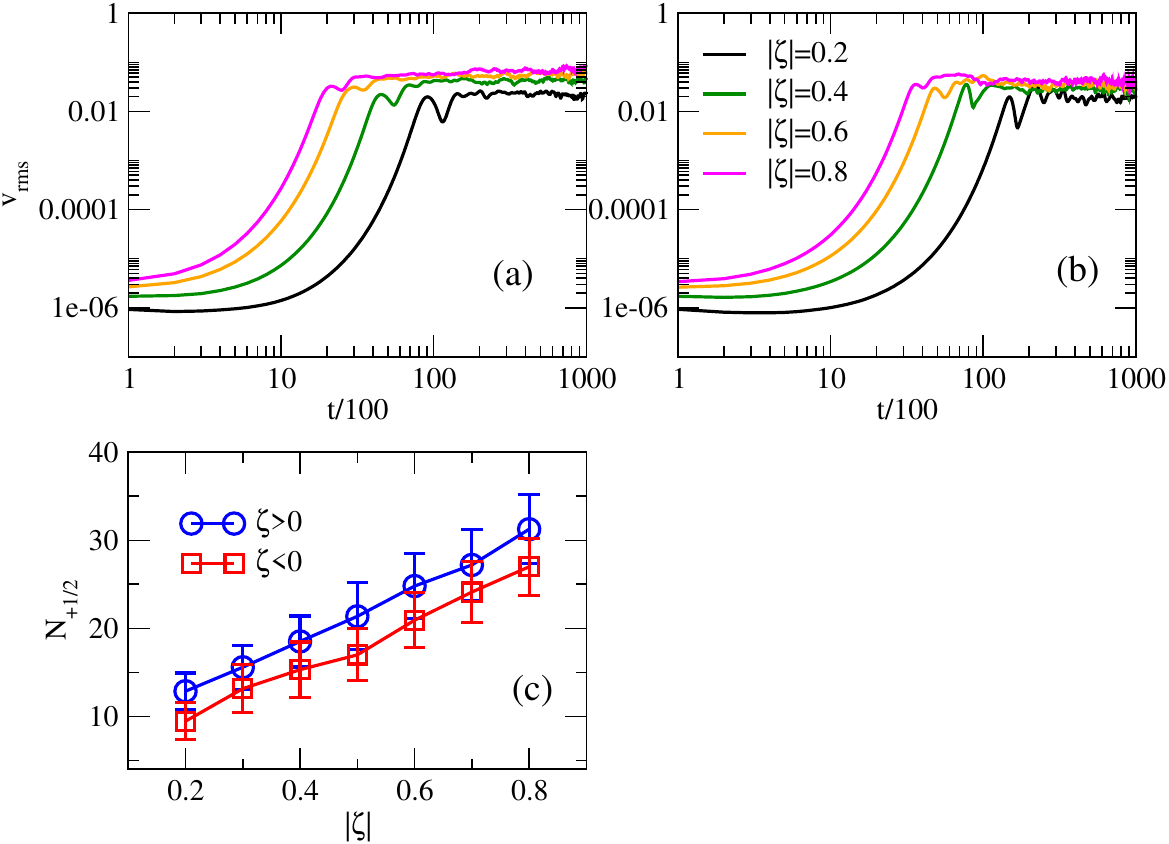}}  
\caption{Root mean square velocity, $v_{rms}$  vs. time $t$ for different strengths of activity, $\zeta>0$ (a) and $\zeta<0$ (b).  \mk{Variations of the }{n}umber of $+1/2$ defects, $N_{+1/2}$ (nematic) \mk{with activity $\vert \zeta \vert $ are shown} for $\zeta>0$ and $\zeta<0$. }
\label{fig:116}
\end{figure}

We start with a uniform initial configuration (where the local nematic directors are aligned in the x-direction). We increase the strength of the active stress, $\zeta$ so that the uniformly aligned state shows bend instability and  the
system evolve to the active turbulent state where creation and annihilation of $\pm 1/2$ defects occur. {Achieving} the steady state \mk{is validated} by calculating the root mean square velocity, $v_{rms}=\sqrt{\bar{v_i^2}}$ and plotting it for extensile (Fig. \ref{fig:116}(a)) and contractile  (Fig. \ref{fig:116}(b)) active stresses \mk{at varied activity} strength, $\vert \zeta \vert$. We also \mk{show} the \mk{variation of the number of $+1/2$ nematic defects with} active strength, $\vert \zeta \vert$  in Fig. \ref{fig:116}(c). These plots \mk{reveal} that as  the activity strength \mk{increases,} the active turbulent state \mk{is} achieved at \mk{a} faster rate \mk{,} resulting in \mk{an} increase \mk{in} the number of topological defects.   

\end{document}